\documentclass[amsmath, amsthm, amssymb, aps, prb, superscriptaddress, twocolumn, nofootinbib, 10pt]{revtex4-1}

\makeatletter
\def\@seccntformat#1{\csname the#1\endcsname.\quad}
\renewcommand\section{\@startsection {section}{1}{0pt}%
  {-3.5ex \@plus -1ex \@minus -.2ex}%
  {2.3ex \@plus.2ex}%
  {\normalfont\normalsize\bfseries\raggedright}}
\renewcommand\subsection{\@startsection {subsection}{2}{0pt}%
  {-3.25ex\@plus -1ex \@minus -.2ex}%
  {1.5ex \@plus .2ex}%
  {\normalfont\normalsize\bfseries\raggedright}}
\makeatother



\usepackage[colorlinks=true,linkcolor=blue,citecolor=blue,urlcolor=blue]{hyperref}
\usepackage{amsmath,amssymb,amsfonts}
\usepackage{graphicx}
\usepackage{cleveref}
\crefname{figure}{Fig.}{Figs.}

\usepackage{siunitx}
\usepackage[resetlabels,labeled]{multibib}
\usepackage{etoolbox}
\usepackage{float}
\usepackage{geometry}
\usepackage{natbib}
\usepackage{mathrsfs}
\usepackage{textcomp}
\usepackage{lipsum}
\usepackage{xcolor}
\usepackage{upgreek}
\usepackage{gensymb}
\usepackage{siunitx}

\begin{document}

\preprint{APS/123-QED}

\title{Micro-Transfer Printed Continuous-Wave and Mode-Locked Laser Integration at 800 nm on a Silicon Nitride Platform}




\author{Max~Kiewiet$^{1,2*}$, Stijn~Cuyvers$^{1,2}$, Maximilien~Billet$^{1,2}$, Konstantinos~Akritidis$^{1,2}$, Valeria~Bonito~Oliva$^{1,2}$, Gaudhaman~Jeevanandam$^{2}$, 
Sandeep~Saseendran$^{2}$, Manuel~Reza$^{2}$, Pol~Van~Dorpe$^{2}$, Roelof~Jansen$^{2}$, Joost~Brouckaert$^{2}$, G\"unther~Roelkens$^{1,2}$, Kasper~Van~Gasse$^{1,2}$ and Bart~Kuyken$^{1,2,\dagger}$\\ 
\vspace{+0.1 in}
\textit{\small{
$^1$Photonics Research Group, INTEC, Ghent University - imec, 9052 Ghent, Belgium\\
$^2$imec, Kapeldreef 75, 3001 Leuven, Belgium. \\
{\small $^*$Max.Kiewiet@ugent.be, $^\dagger$Bart.Kuyken@ugent.be}}}}

\date{\today}

\begin{abstract}
Applications such as augmented and virtual reality (AR/VR), optical atomic clocks, and quantum computing require photonic integration of (near-)visible laser sources to enable commercialization at scale. The heterogeneous integration of III-V optical gain materials with low-loss silicon nitride waveguides enables complex photonic circuits with low-noise lasers on a single chip. Previous such demonstrations are mostly geared towards telecommunication wavelengths. At shorter wavelengths, limited options exist for efficient light coupling between III-V and silicon nitride waveguides. Recent advances in wafer-bonded devices at these wavelengths require complex coupling structures and suffer from poor heat dissipation. Here, we overcome these challenges and demonstrate a wafer-scale micro-transfer printing method integrating functional III-V devices directly onto the silicon substrate of a commercial silicon nitride platform. We show butt-coupling of efficient GaAs-based amplifiers operating at 800 nm with integrated saturable absorbers to silicon nitride cavities. This resulted in extended-cavity continuous-wave and mode-locked lasers generating pulse trains with repetition rates ranging from 3.2 to 9.2 GHz and excellent passive stability with a fundamental radio-frequency linewidth of 519 Hz. These results show the potential to build complex, high-performance fully-integrated laser systems at 800 nm using scalable manufacturing, promising advances for AR/VR, nonlinear photonics, timekeeping, quantum computing, and beyond.
\end{abstract}

\maketitle
\mbox{}
\clearpage
\begin{figure*}[t!]
    \centering
    \includegraphics[width=\textwidth]{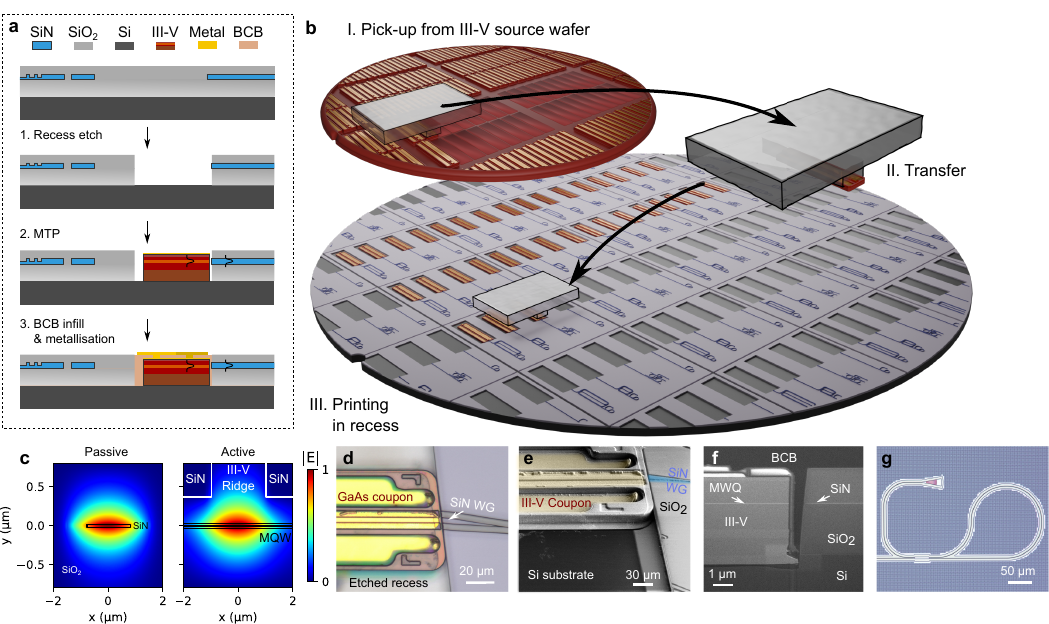}
    \caption{\textbf{Heterogeneous integration platform on silicon nitride with different laser structures at a wavelength centered at 800 nm.} \textbf{a.} Simplified process flow for the micro-transfer printing process. Steps shown: 1. Recess etch in top-cladded SiN die; 2. MTP in recess where III-V mode is vertically aligned with the SiN mode, which are indicated in the figure. 3. Post-printing processing: planarization and infill with benzocyclobutene (BCB), electrical via etching, and contact pad metallization. \textbf{b.} Illustration of the proposed wafer-scale micro-transfer printing approach showing (I). The pick-up from the III-V source wafer using an elastomer stamp; (II). Transfer to the target wafer; (III). Printing in recess etched on target wafer. \textbf{c.} Optical mode comparison between passive SiN mode in a shallow etched taper of \SI{3}{\micro\meter} $\times$ \SI{60}{\nano\meter} with a calculated overlap of 96 \%. \textbf{d.} Microscope image of a micro-transfer printed coupon aligned to a SiN waveguide showing excellent alignment. \textbf{e.} SEM image (false colored) showing the same coupon in the recess before BCB infill. The position of the SiN waveguide (SiN WG) is visible due to superficial damage on the silicon oxide top cladding caused by the recess etch due to the topography of the SiN waveguide. \textbf{f.} SEM image of focused ion-beam-etched cross-section of a printed coupon. A facet distance of less than 500 nm is visible, limited by the angle of the recess etch. \textbf{g.} Microscope image of silicon nitride waveguide components: a grating coupler and a Sagnac loop mirror using an asymmetric multi-mode interferometer (MMI).}
    \label{fig:mainfig}
\end{figure*}

\begin{figure*}[t]
    \centering
    \includegraphics[width=\textwidth]{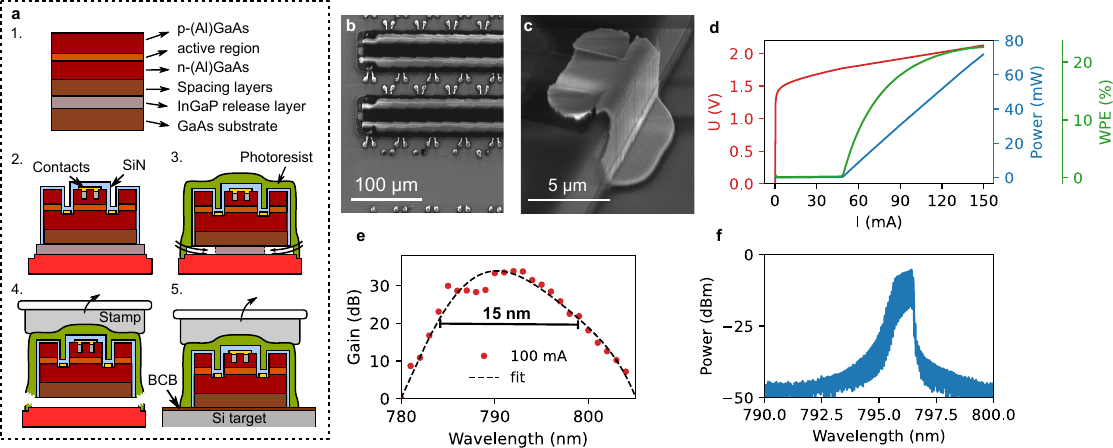}
    \caption{\textbf{Process steps for coupon preparation for micro-transfer printing and characterization on the silicon target chip} \textbf{a.} Schematic process flow for coupon preparation and the transfer to silicon. Steps shown: 1. Epitaxial layer stack featuring waveguide layers, spacing layers to align III-V mode with SiN mode and release layer to suspend coupons; 2. Coupons are patterned featuring a \SI{2}{\micro\meter}-wide ridge waveguide, etched facets, SiN passivation, and metal contacts; 3. Photoresist encapsulation anchoring the coupons to the substrate and selective under-etch to suspend the coupons; 4. Pick-up using elastomer stamp, breaking the photoresist tethers. 5. Printing on a silicon substrate using a thin $\sim$ \SI{50}{\nano\meter} BCB adhesion layer. \textbf{b.} SEM image of source substrate showing suspended coupons and one picked coupon. \textbf{c.} Detail of coupon showing backside highy-reflective (HR) gold coating. \textbf{d.} Continuous-wave (CW) LIV characteristics of a transfer-printed FP laser on silicon including wall-plug efficiency (WPE), measured using a flat free-space silicon power meter. \textbf{e.} Internal small-signal gain excluding the 6 dB fiber coupling loss versus wavelength of a printed RSOA on silicon. \textbf{f.} Optical emission spectrum of a printed FP laser on silicon, measured with a \SI{30}{\pico\meter} resolution bandwidth.}
    \label{fig:MTPonSi}
\end{figure*}

\begin{figure*}[t]
    \centering
    \includegraphics[]{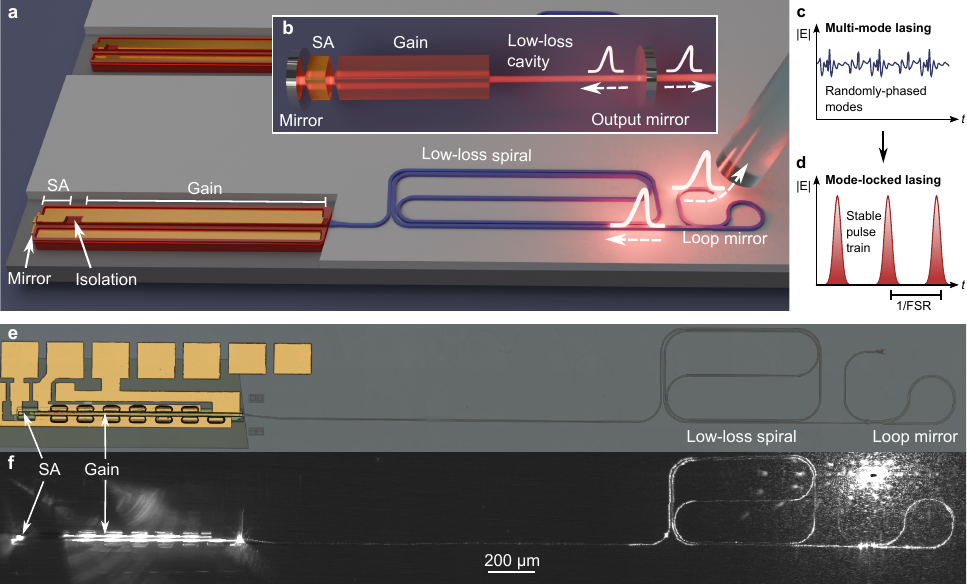}
    \caption{\textbf{Schematic and the properties of the integrated mode-locked laser} \textbf{a.} Illustration of a fully integrated extended-cavity mode-locked laser on silicon nitride. \textbf{b.} (inset) Equivalent free-space mode-locked laser system. \textbf{c.} Schematic time-dependent output field of Fabry-Perot laser without mode-locking. \textbf{d.} Schematic time-dependent output field of a mode-locked laser with a saturable absorber. \textbf{e.} Bright-field microscope image of a mode-locked laser. \textbf{f.} Dark-field microscope image of the same laser under 50 mA gain current with forward-biased SA, showing showing side-wall scattering and emission from the grating coupler.}
    \label{fig:mllexplanation}
\end{figure*}

\begin{figure*}[t]
    \centering
    \includegraphics[]{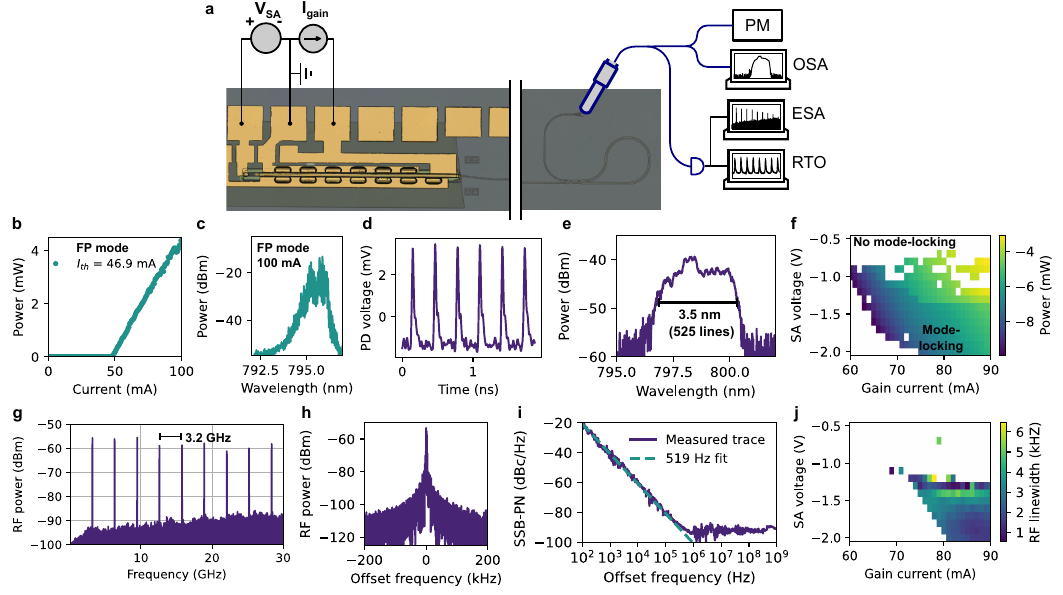}
    \caption{\textbf{Performance of the extended-cavity continuous-wave and mode-locked lasers} \textbf{a.} Measurement setup with PM: power meter; OSA: optical spectrum analyzer; ESA: electrical spectrum analyzer; RTO: real-time oscilloscope. \textbf{b.}  CW LI curve of a 9.2 GHz FSR extended-cavity Fabry-Perot laser. \textbf{c.} Optical FP spectrum at 100 mA gain current measured with a resolution bandwidth of 30 pm. \textbf{d.} Pulse train of a 3.2 GHz MLL at 85 mA gain current and -1.3 V SA bias as measured on a 25 GHz photodiode. \textbf{e.} Widest achieved mode-locked optical spectrum at 90 mA gain current and -1.3 V SA bias measured with a resolution bandwidth of 30 pm. \textbf{f.} Mode-locking map showing average waveguide-coupled output power for different gain currents and saturable absorber bias voltages, where white data points correspond to laser operation modes without mode-locking, as determined from the RF spectrum. \textbf{g.} RF comb at optimal mode-locking point of 85 mA gain current and -1.3 V SA bias. for a resolution bandwidth (RBW) of 100 kHz \textbf{h.} Fundamental RF line, measured with RBW of 100 Hz. \textbf{i.} Single sideband phase noise (SSB-PN) measurement of fundamental RF line along with a 519 Hz Lorentzian fit. \textbf{j.} Mode-locking map of fundamental Lorentzian RF linewidth as fitted from SSB-PN measurement, where white data points correspond to laser operation modes without mode-locking which was stable enough for a SSB-PN measurement.}
    \label{fig:MLLperformance}
\end{figure*}


\section{\label{sec:introduction}Introduction}
Miniaturized, energy-efficient optical devices are increasingly crucial for a wide range of technologies, from advanced sensors and communication systems to quantum technologies and precision timekeeping. Photonic integrated circuits (PICs), which combine multiple optical functions on a single chip, leverage existing CMOS (Complementary Metal-Oxide-Semiconductor) processes and infrastructure to achieve this miniaturization while offering significant advantages in terms of size, weight, power, and cost (SWaP-C) compared to traditional optical systems \cite{shekhar2024roadmapping, porcel2019silicon}. \\
To realize advanced PICs for applications beyond telecommunication such as precision timekeeping with integrated atomic clocks, sensing, and quantum computing with rubidium (Rb) atoms, integration of laser sources providing coherent radiation at shorter wavelengths on a waveguide platform suitable for this wavelength is required \cite{newman2019architecture, mehta2020integrated,niffenegger2020integrated,zinoviev2011integrated,isichenko2023photonic}. For wavelengths shorter than \SI{1.1}{\micro\meter}, the widely used silicon-on-insulator (SOI) platform is no longer suitable due to its relatively narrow band-gap \cite{thomson2016roadmap}. The most promising and technologically mature solution to this problem is the use of a silicon nitride (SiN) waveguide platform, offering broadband transparency down to a wavelength of approximately 400 nm and extremely low losses of less than 0.1 $\mathrm{dB\ m^{-1}}$ \cite{blumenthal2018silicon,Xiang:22}. This makes SiN waveguides especially attractive for nonlinear devices leveraging ultra-high-Q microcavities or other co-integrated nonlinear materials \cite{puckett2021422}. \\
However, a crucial challenge lies in the coupling of light from the III-V light source to the low-loss waveguide due to the large index contrast between III-V materials ($>3$) and SiN ($\sim2$). Many promising demonstrations of III-V on SiN have been shown at telecommunication and optical communication wavelengths (1550~nm and 1310~nm), using evanescent tapers and intermediate thin silicon layers \cite{cuyvers2021low, op2020heterogeneous, xiang2020narrow, xiang2021high, zhang2018transfer, zhang2023iii}. Unfortunately, this is not possible at shorter wavelengths due to the small band gap of silicon.\\
Using an intermediate dielectric as a coupling structure/spot-size converter, light can be non-adiabatically coupled from the III-V mode to the intermediate dielectric waveguide and subsequently evanescently coupled to the SiN waveguide using an inverted taper \cite{roelkens2006laser}. This was successfully used to integrate a broad range of III-V functionalities on a SiN platform first at sub-micrometer wavelengths \cite{tran2022extending} and subsequently at 780 nm demonstrating rubidium spectroscopy using wafer-bonding with high waveguide-coupled powers up to 12 mW \cite{zhang2023photonic, castro2025integrated}. However, such dielectric coupling structures require the deposition and patterning of waveguiding structures after III-V integration, complicating the post-integration process flow significantly. Furthermore, the use of adiabatic inverted tapers limits the operating bandwidth.\\
An alternative light coupling approach, butt-coupling, has widely been used to integrate well-performing lasers on SiN and thin-film lithium niobate (TFLN) from 1550~nm all the way down to 637~nm to make both stable single-mode \cite{fan2016optically, fan2020hybrid, winkler2024widely, boller2020hybrid} and mode-locked lasers \cite{klaver2021self, winkler2024chip, vissers2021hybrid, vissers2022hybrid}. Here, a III-V gain chip is butt-coupled to a chip with low-loss waveguides. Due to the die-level nature of this process, chip-to-chip coupling is inherently less scalable than wafer-scale methods, limiting its commercialization potential.\\
Through the superior flexibility of the micro-transfer printing integration method \cite{roelkens2024present}, III-V can be heterogeneously integrated at a wafer scale using the direct butt-coupling approach, where III-V is placed directly on top of the silicon substrate, offering superior thermal characteristics compared to wafer bonding integration techniques \cite{justice2012wafer,loi2018thermal} as well as extremely broad-band light coupling. Furthermore, micro-transfer printing offers further advantages such as efficient III-V material use, the potential of using pre-characterized lasers and SOAs, and no required III-V processing on the SiN wafer. This technique was previously demonstrated by coupling Fabry-Perot lasers to a waveguide at 1550 and 1310 nm wavelengths \cite{juvert2018integration, uzun2023integration}. However, these demonstrations lack efficient coupling and more complex laser functionality such as extended-cavity lasers.\\
This work features the first extended-cavity laser demonstration of wafer-scale compatible butt-coupled integration using micro-transfer printing at short wavelengths. This is used to achieve lasers emitting at 800 nm, leveraging the wavelength-agnostic nature of butt-coupling. We demonstrate extended-cavity lasers with more than 4 mW continuous-wave (CW) output power and generation of ultra-stable pulse trains through passive mode-locking with fundamental radio-frequency (RF) linewidths down to 519 Hz and pulse energies up to 0.27 pJ. This is achieved by leveraging our highly efficient micro-transfer printed GaAs-based laser 'coupons', which show laser power exceeding 70 mW and semiconductor optical amplifier (SOA) internal gain of more than 30 dB. The integrated lasers demonstrate the robustness and versatility of our integration approach and can serve as key enablers for fully integrated photonics at 800 nm for spectroscopy, microwave photonics, quantum computing, and chip-scale optical atomic clocks \cite{coddington2016dual, mcferran2005low, telle1999carrier, marpaung2019integrated}.

\section{\label{sec:results} Results}
\subsection{\label{sec:platform} III-V on S\MakeLowercase{i}N photonics platform through micro-transfer printing}
This work proposes a platform with heterogeneously integrated III-V/SiN devices, where \SI{5}{\micro\meter}-thick III-V epitaxial layers are micro-transfer printed in an etched recess in a SiN waveguide platform. A simplified integration process flow is illustrated in Fig.~\ref{fig:mainfig}a. A more detailed description is provided in the methods section and the supplementary material. The integration process is done at die-level, however, micro-transfer printing is a fully wafer-scale compatible process, meaning the full integration process can also be executed at the wafer-level. Fig.~\ref{fig:mainfig}b shows an outlook on the micro-transfer printing process at the wafer-level. Here, functional pre-processed III-V devices such as lasers are picked up from the III-V source wafer and transferred to the SiN target wafer where they are placed directly on the Si substrate in an etched recess, butt-coupled to the SiN waveguide. The modes in the passive SiN waveguide and the active III-V waveguide are compared in Fig.~\ref{fig:mainfig}c, showing a 96\% overlap of the modal shapes. In Fig.~\ref{fig:mainfig}d-f, the micro-transfer printed coupon on the SiN platform is shown along with the alignment to the SiN waveguide. Fig.~\ref{fig:mainfig}g shows the Sagnac loop mirror used in all laser structures.
To couple light between III-V and SiN waveguides, the optical modes are vertically aligned by tuning the epitaxial layer stack thicknesses to match the recess depth etched in the SiN waveguide platform. A more detailed description of this process is provided in the supplementary material. This process requires excellent alignment accuracy in three dimensions: the distance between III-V and SiN facets, the lateral printing offset due to MTP inaccuracy, and the vertical misalignment due to inaccurate buried oxide and epitaxial layer thicknesses. As can be seen in Fig.~\ref{fig:mainfig}f, the facet distance is limited by the angle of the recess etch. This angle is measured to be approximately 85 degrees, resulting in a minimum facet distance of approximately 450 nm. From simulations shown in the supplementary material, the coupling efficiencies for horizontal misalignment values within tool specifications of $\pm$\SI{1}{\micro\meter} $(3\sigma)$ exceed 50\% with a maximum of 86\%. Better coupling can be achieved by using a more advanced, wafer-scale micro-transfer printing tool with alignment specifications of $\pm$\SI{0.5}{\micro\meter} $(3\sigma)$.\\

\subsection{\label{sec:silicon} High-Power III-V laser coupons for butt-coupled Micro-Transfer Printing}
The simplified process flow for source wafer preparation for micro-transfer printing of (Al)GaAs-based laser coupons is illustrated in Fig.~\ref{fig:MTPonSi}a., starting from epitaxially grown III-V layers featuring four quantum wells, along with the printing on a dummy silicon substrate for testing of the coupons. In Fig.~\ref{fig:MTPonSi}b, suspended laser coupons on the GaAs source substrate are shown. Since the direct-butt coupling scheme allows for coupling at only one facet, the rear facet of the coupons is coated with a highly-reflective gold coating, as shown in Fig.~\ref{fig:MTPonSi}c. Both reflective SOA (RSOA) and Fabry-Perot coupons are fabricated on the same source wafer, where the only difference is the etched front facet angle. To reduce reflections into the RSOA waveguide mode from a simulated value of -12 dB down to -70 dB (as shown in the supplementary material), the front facet is angled 7 degrees with respect to the waveguide direction. The Fabry-Perot coupons have a 0-degree facet angle which provides a reflection with a simulated value of 6\% into the waveguide mode (when encapsulated in BCB) to facilitate lasing.\\
The light-current-voltage (LIV) characteristics of a typical 1 mm-long FP coupon are shown in \ref{fig:MTPonSi}d, measured using a free-space silicon power meter. The FP lasers show powerful lasing with output powers exceeding 70 mW at 150 mA and a threshold of 48 mA or 2.4 $\mathrm{kAcm^{-2}}$. Furthermore, the excellent heat-sinking characteristics of this integration approach are apparent in the lack of thermal roll-off even at the highest measured current densities exceeding $\mathrm{7\ kAcm^{-2}}$. The typical high efficiency of GaAs-based quantum well lasers is evident from the wall-plug efficiencies exceeding 20\% and the slope efficiency of 0.66 $\mathrm{WA^{-1}}$. The corresponding optical spectrum at 100 mA is shown in Fig.~\ref{fig:MTPonSi}f, showing emission at 796 nm. Lastly, using a lensed fiber to couple into the III-V waveguide mode, the small-signal gain of a 1 mm-long RSOA is extracted and plotted in Fig.~\ref{fig:MTPonSi}e. Here, the 6 dB fiber-to-III-V coupling efficiency is compensated to show a high internal gain exceeding 30 dB and a 20 dB gain over a wide bandwidth of 15 nm. This shows the potential of the RSOA to be used in widely tunable single-mode Vernier lasers for rubidium spectroscopy.

\subsection{\label{sec:MLLs} Heterogeneously integrated continuous-wave and mode-locked lasers at 800 nm}
For nonlinear applications such as octave-spanning comb generation for atomic clocks, high optical powers are required. As a consequence of the nonlinearity of such effects, sources generating short pulses can be used to relax the requirement on average laser power. A mode-locked laser generates pulses fully on-chip by using a saturable absorber (SA) to lock the phases of different longitudinal cavity modes \cite{keller2021ultrafast}. Such an on-chip mode-locked laser is illustrated in Fig.~\ref{fig:mllexplanation}a, along with an equivalent free-space laser system in Fig.~\ref{fig:mllexplanation}b. Here, a saturable absorber is placed in a cavity otherwise consisting of two mirrors and an optical gain element. By adding the saturable absorber, the initially randomly phased longitudinal modes of the Fabry-Perot cavity (Fig.~\ref{fig:mllexplanation}c) are phase-locked to form pulses (Fig.~\ref{fig:mllexplanation}d) \cite{571743}. 
In the chip-integrated laser, the cavity is formed using a III-V RSOA coupon with a gold mirror and a 75\% reflecting Sagnac loop mirror. By electrically isolating a subsection of the SOA from the rest of the gain section, a (negative) voltage can be applied to it to cause absorption, while the much longer gain section is positively biased to provide net gain. In the SA, at high incident power, electrons accumulate in the conduction band, depleting the ground state and occupying the excited states. This causes bleaching of the absorption and therefore saturation, allowing for mode-locked operation. \\
To achieve low-noise operation of the laser cavity, the photon lifetime should be maximized \cite{paschotta2004noise, paschotta2006optical}. This is accomplished using a low-loss waveguide spiral in SiN, extending the photon lifetime far beyond what is possible in monolithic III-V platforms \cite{komljenovic2015widely, van2020ring}. Extending the cavity also allows the generation of lower-repetition frequency combs, increasing the spectral density of the optical comb, which is of interest in spectroscopic applications. The fully integrated mode-locked laser is shown in a bright-field microscope image in Fig.~\ref{fig:mllexplanation}e and in a dark-field image slightly above laser threshold in \ref{fig:mllexplanation}f, clearly showing extended-cavity lasing, as evident from the grating coupler emission and side-wall scattering. \\
Using the proposed integration process, multiple mode-locked laser devices are fabricated with free-spectral ranges of 3.2 GHz, 7.5 GHz, and 9.2 GHz by varying the length of the low-loss SiN waveguide spiral in the laser cavity. The 3.2 GHz and 9.2 GHz lasers are made on the imec 200 mm SiN platform, and the 7.5 GHz laser is fabricated on an in-house electron-beam lithography (EBL) platform. The lasers are tested in both the continuous-wave Fabry-Perot mode, where the isolated gain section is forward biased in parallel with the gain section to provide maximum output power, and the mode-locked mode, where the isolated gain section is used as a saturable absorber by biasing it separately to cause mode-locking. \\
The measurement setup is illustrated in Fig.~\ref{fig:MLLperformance}a in the mode-locking operation mode. The LI performance of the 9.2 GHz imec platform laser in CW operation mode is shown in Fig.~\ref{fig:MLLperformance}b. The corresponding spectrum is plotted in Fig.~\ref{fig:MLLperformance}c. at 100 mA gain current, showing multi-mode lasing. In this work, we show a fully-integrated butt-coupled extended-cavity laser with waveguide-coupled output powers exceeding 4 mW and a threshold of 48 mA or 2.4 $\mathrm{kAcm^{-2}}$. The current coupling efficiency is limited by the control of the recess etch and slight bending in the coupon due to adhesion issues resulting from the prototyping stage of the MTP development. Optimization of the coupling can be done to unlock the power levels shown in Fig.~\ref{fig:MTPonSi}. From the coupling simulations, we expect to achieve waveguide-coupled Fabry-Perot laser output power up to 50 mW after optimization. \\
Stable mode-locking was observed for all three repetition rate mode-locked lasers, which are compared in the supplementary material. The mode-locking is characterized in more detail in Fig.~\ref{fig:MLLperformance}b-j for the 3.2 GHz imec platform laser. A mode-locking map is measured for this laser by performing a two-dimensional sweep of the SA voltage and gain current. This is shown in Fig.~\ref{fig:MLLperformance}f and j, where the average output power and RF linewidth are plotted as a function of the two input parameters for only those combinations where mode-locking is observed, defined as showing a stable RF comb with at least three equidistant RF tones. Pulse energies up to 0.16 pJ are observed for the 3.2 GHz MLL and up to 0.27 pJ for the 7.5 GHz MLL. These powers are, as described previously, limited by the III-V to SiN coupling. With optimized coupling, we project pulse energies in the order of 3 pJ. The optimal mode-locking point at 85 mA gain current and -1.3 V SA bias is further characterized in Fig.~\ref{fig:MLLperformance}d and e, showing a real-time oscilloscope trace of the photodiode voltage to show pulsed operation and the optical spectrum which exhibits a wide, flat emission spectrum with a 3.5 nm 10-dB bandwidth, corresponding to 1.7 THz and containing 525 comb lines. In Fig.~\ref{fig:MLLperformance}g and h, the optimal mode-locking point is further characterized using the RF comb and a zoom on the fundamental RF tone, showing an extinction ratio of approximately 50 dB, limited by the noise floor of the measurement. Furthermore, from the corresponding single-sideband phase-noise measurement in Fig.~\ref{fig:MLLperformance}i, a Lorentzian linewidth of 519 Hz is extracted, corresponding to a minimum pulse-to-pulse timing jitter of 51 fs indicating excellent passive stability of the mode-locked comb. A more detailed characterization of the mode-locked lasers is found in the supplementary materials. These results show the potential of the outlined integration method to create powerful and complex laser systems for nonlinear applications.\\

\section{\label{sec:discussion}Discussion}
Using the integration method outlined in this work, complex and high-power lasers can be integrated on a SiN platform for applications at 800 nm. The output power of the fabricated devices can be improved to match the high-power potential seen in the characterization of the laser coupons by optimizing the III-V to SiN coupling. This can be improved significantly by optimizing the micro-transfer printing process by tuning the printing parameters, the adhesion layer composition and thickness, and the etch depth control of the recess etch. Furthermore, single-mode lasers can be integrated by including filters such as distributed Bragg reflectors or Vernier ring filters. The latter can be integrated to potentially achieve $>15$ nm wavelength tunability. Lastly, optimal potential power can be achieved by integrating Fabry-Perot laser coupons with flat front facets. In this way, waveguide-coupled powers of up to 50 mW could potentially be attained.\\
The demonstrated integration method can be readily extended to even shorter wavelengths using different gain materials such as InGaP, (Al)GaN, and AlN, owing to the wavelength-agnostic nature of the coupling scheme. Furthermore, different waveguide platforms such as thin-film lithium-niobate-on-insulator (LNOI) or aluminium oxide can be used to further extend the spectrum and functionalities. By using the flexibility of the micro-transfer printing approach, multiple different materials can be co-integrated at a density and complexity largely unattainable using wafer bonding. Combining lasers from this work with micro-transfer printed evanescently-coupled lithium-niobate or lithium-tantalate traveling-wave modulators \cite{vanackere2023heterogeneous, niels2025high}, high-speed optical interconnects or fast-tunable single-mode lasers can be created at 800 nm. Furthermore, silicon photodetectors can be co-integrated as monitor diodes or as sensing diodes for on-chip spectrometers or absolute frequency stabilization using rubidium gas cells. Combining a mode-locked laser demonstrated in this work with a high-Q SiN ring resonator or a micro-transfer printed nonlinear GaP waveguide \cite{billet2022gallium}, fully integrated and efficient supercontinuum sources can be achieved for f-2f referencing in chip-integrated atomic clocks.
Finally, the compatibility of this integration process with existing photonic SiN platforms means this approach is well-suited for larger scale high-volume manufacturing. Micro-transfer printing can be used to integrate III-V material at a wafer scale on the back-end-of-line of a CMOS process, keeping CMOS-incompatible III-V material out of the front-end. The efficient use of expensive III-V materials and the demonstrated flexibility exemplify the potential of the micro-transfer printing technique for III-V integration at shorter wavelengths.

\section{Methods}
\subsection*{III-V laser coupon fabrication}
The laser coupon fabrication starts with epitaxial layers grown using metal-organic vapor phase epitaxy (MOVPE) on a 2-inch n-doped GaAs substrate, shown in more detail in the supplementary material. The laser ridge waveguide and saturable absorber isolation are formed with BCl\textsubscript{3}/H\textsubscript{2}-based inductively-coupled plasma (ICP) etching using a SiN hard-mask deposited using plasma-enhanced chemical vapor deposition (PECVD) and patterned through UV lithography and SF\textsubscript{6}/CF\textsubscript{4}/H\textsubscript{2}-based reactive-ion etching (RIE). Next, the laser is passivated by low-stress PECVD SiN, vias are opened, and P-contact metal (Ti/Au) and N-contact metal (Ni/Ge/Au) are deposited using electron-beam deposition and a lift-off process. To eliminate native oxide at the metal/III-V interface, a dilute HCl dip is used prior to contact deposition and the contacts are rapid-thermal annealed (RTA) at 430 \textdegree C. Subsequently, the mesa of the laser, including the optical facets, is etched using the same ICP recipe and passivated using PECVD SiN. A gold mirror is deposited on one facet, including a thin (a few nm-thick to limit absorption) Ti adhesion layer, using angled electron-beam deposition and a lift-off process. Next, the InGaP release layer is patterned using BCl\textsubscript{3}/H\textsubscript{2}-based ICP etching and a photoresist mask to selectively expose the substrate. The coupons are then encapsulated with thick ($\mathrm{\sim}$\SI{6}{\micro\meter}) positive photoresist forming tethers to the substrate and leaving the front facet bare to allow close butt-coupling. Finally, the coupons are under-etched using a 2:1 HCl:H\textsubscript{2}O solution.
\subsection*{III-V coupon characterization on silicon substrate}
The FP laser coupons transferred to a Si substrate are electrically probed using DC probes and characterized using a flat free-space photodiode (Thorlabs S130C) and an OSA (Anritsu MS9740A). The gain of the RSOA coupons is extracted using the a lensed fiber for coupling and a tunable titanium-sapphire laser. This is done by measuring the reflected and amplified laser output using the same OSA after a circulator and dividing by the input laser peak, at different wavelengths. The fiber-to-RSOA coupling is extracted by fitting the fiber-coupled LI curve to the free-space LI curve of the amplified spontaneous emission of the RSOA.
\subsection*{Micro-transfer printing integration}
The III-V coupons are integrated on both the 200 mm SiN platform of imec and an in-house EBL platform. For the imec platform, a 200 mm wafer is provided featuring 300 nm-thick SiN waveguides with silicon oxide (SiO\textsubscript{2}) top and bottom claddings. For the in-house platform, uniform wafers with a 300 nm-thick SiN layer on 3300 nm SiO\textsubscript{2} on Si substrate are patterned using EBL and RIE etching in CHF\textsubscript{3}-based chemistry before adding a \SI{2}{\micro\meter}-thick top cladding using inductively-coupled plasma chemical vapor deposition (ICP-CVD). Before micro-transfer printing, a recess is etched using ICP and a CHF\textsubscript{3}/Ar gas mixture and a chromium metal hard-mask, patterned using a lift-off process. Next, a thin ($\mathrm{\sim 50\ nm}$) adhesion layer of photo-patternable BCB (Cyclotene 4000 series) is added. This is patterned using UV lithography to remove any build-up at the recess edges. Next, the III-V coupons are micro-transfer printed into the recess, aligned to the silicon nitride waveguide. Lastly, post-printing processing is done to add metal contacts and fill up any voids in the optical path with BCB. The integration process is described in more detail in the supplementary material. 
\subsection*{Extended-cavity laser characterization}
The extended-cavity CW and mode-locked lasers are characterized using the output grating coupler and a cleaved and AR-coated 780HP single-mode fiber. Using fiber-based splitters, the optical output of the mode-locked lasers is measured using a fiber-coupled power meter (HP 1936-R) and an OSA (Anritsu MS9740A) using a resolution of 0.03 nm. The waveguide-coupled power is calculated using the grating coupler insertion loss extracted with use of a cut-back structure. Furthermore, the optical signal is measured using a 25 GHz bandwidth GaAs-based photodiode (Thorlabs DXM25CF). The resulting RF signal is analyzed with a 63 GHz bandwidth RTO (Keysight DSAZ634A), and a 44 GHz bandwidth ESA (Keysight N-9010A). The phase-noise measurement is taken using the ESA. These measurements are repeated for different gain currents and SA voltages to map the resulting figures of merit in the two-dimensional parameter space.

\bibliography{bibliography}

\providecommand{\noopsort}[1]{}\providecommand{\singleletter}[1]{#1}%
\begin{thebibliography}{10}
\expandafter\ifx\csname url\endcsname\relax
  \def\url#1{\texttt{#1}}\fi
\expandafter\ifx\csname urlprefix\endcsname\relax\def\urlprefix{URL }\fi
\providecommand{\bibinfo}[2]{#2}
\providecommand{\eprint}[2][]{\url{#2}}

\bibitem{shekhar2024roadmapping}
\bibinfo{author}{Shekhar, S.} \emph{et~al.}
\newblock \bibinfo{title}{Roadmapping the next generation of silicon photonics}.
\newblock \emph{\bibinfo{journal}{Nature Communications}} \textbf{\bibinfo{volume}{15}}, \bibinfo{pages}{751} (\bibinfo{year}{2024}).

\bibitem{porcel2019silicon}
\bibinfo{author}{Porcel, M.~A.} \emph{et~al.}
\newblock \bibinfo{title}{Silicon nitride photonic integration for visible light applications}.
\newblock \emph{\bibinfo{journal}{Optics \& Laser Technology}} \textbf{\bibinfo{volume}{112}}, \bibinfo{pages}{299--306} (\bibinfo{year}{2019}).

\bibitem{newman2019architecture}
\bibinfo{author}{Newman, Z.~L.} \emph{et~al.}
\newblock \bibinfo{title}{Architecture for the photonic integration of an optical atomic clock}.
\newblock \emph{\bibinfo{journal}{Optica}} \textbf{\bibinfo{volume}{6}}, \bibinfo{pages}{680--685} (\bibinfo{year}{2019}).

\bibitem{mehta2020integrated}
\bibinfo{author}{Mehta, K.~K.} \emph{et~al.}
\newblock \bibinfo{title}{Integrated optical multi-ion quantum logic}.
\newblock \emph{\bibinfo{journal}{Nature}} \textbf{\bibinfo{volume}{586}}, \bibinfo{pages}{533--537} (\bibinfo{year}{2020}).

\bibitem{niffenegger2020integrated}
\bibinfo{author}{Niffenegger, R.~J.} \emph{et~al.}
\newblock \bibinfo{title}{Integrated multi-wavelength control of an ion qubit}.
\newblock \emph{\bibinfo{journal}{Nature}} \textbf{\bibinfo{volume}{586}}, \bibinfo{pages}{538--542} (\bibinfo{year}{2020}).

\bibitem{zinoviev2011integrated}
\bibinfo{author}{Zinoviev, K.~E.}, \bibinfo{author}{Gonz{\'a}lez-Guerrero, A.~B.}, \bibinfo{author}{Dom{\'\i}nguez, C.} \& \bibinfo{author}{Lechuga, L.~M.}
\newblock \bibinfo{title}{Integrated bimodal waveguide interferometric biosensor for label-free analysis}.
\newblock \emph{\bibinfo{journal}{Journal of lightwave technology}} \textbf{\bibinfo{volume}{29}}, \bibinfo{pages}{1926--1930} (\bibinfo{year}{2011}).

\bibitem{isichenko2023photonic}
\bibinfo{author}{Isichenko, A.} \emph{et~al.}
\newblock \bibinfo{title}{Photonic integrated beam delivery for a rubidium 3d magneto-optical trap}.
\newblock \emph{\bibinfo{journal}{Nature communications}} \textbf{\bibinfo{volume}{14}}, \bibinfo{pages}{3080} (\bibinfo{year}{2023}).

\bibitem{thomson2016roadmap}
\bibinfo{author}{Thomson, D.} \emph{et~al.}
\newblock \bibinfo{title}{Roadmap on silicon photonics}.
\newblock \emph{\bibinfo{journal}{Journal of Optics}} \textbf{\bibinfo{volume}{18}}, \bibinfo{pages}{073003} (\bibinfo{year}{2016}).

\bibitem{blumenthal2018silicon}
\bibinfo{author}{Blumenthal, D.~J.}, \bibinfo{author}{Heideman, R.}, \bibinfo{author}{Geuzebroek, D.}, \bibinfo{author}{Leinse, A.} \& \bibinfo{author}{Roeloffzen, C.}
\newblock \bibinfo{title}{Silicon nitride in silicon photonics}.
\newblock \emph{\bibinfo{journal}{Proceedings of the IEEE}} \textbf{\bibinfo{volume}{106}}, \bibinfo{pages}{2209--2231} (\bibinfo{year}{2018}).

\bibitem{Xiang:22}
\bibinfo{author}{Xiang, C.}, \bibinfo{author}{Jin, W.} \& \bibinfo{author}{Bowers, J.~E.}
\newblock \bibinfo{title}{Silicon nitride passive and active photonic integrated circuits: trends and prospects}.
\newblock \emph{\bibinfo{journal}{Photon. Res.}} \textbf{\bibinfo{volume}{10}}, \bibinfo{pages}{A82--A96} (\bibinfo{year}{2022}).

\bibitem{puckett2021422}
\bibinfo{author}{Puckett, M.~W.} \emph{et~al.}
\newblock \bibinfo{title}{422 million intrinsic quality factor planar integrated all-waveguide resonator with sub-mhz linewidth}.
\newblock \emph{\bibinfo{journal}{Nature communications}} \textbf{\bibinfo{volume}{12}}, \bibinfo{pages}{934} (\bibinfo{year}{2021}).

\bibitem{cuyvers2021low}
\bibinfo{author}{Cuyvers, S.} \emph{et~al.}
\newblock \bibinfo{title}{Low noise heterogeneous iii-v-on-silicon-nitride mode-locked comb laser}.
\newblock \emph{\bibinfo{journal}{Laser \& Photonics Reviews}} \textbf{\bibinfo{volume}{15}}, \bibinfo{pages}{2000485} (\bibinfo{year}{2021}).

\bibitem{op2020heterogeneous}
\bibinfo{author}{Op~de Beeck, C.} \emph{et~al.}
\newblock \bibinfo{title}{Heterogeneous iii-v on silicon nitride amplifiers and lasers via microtransfer printing}.
\newblock \emph{\bibinfo{journal}{Optica}} \textbf{\bibinfo{volume}{7}}, \bibinfo{pages}{386--393} (\bibinfo{year}{2020}).

\bibitem{xiang2020narrow}
\bibinfo{author}{Xiang, C.} \emph{et~al.}
\newblock \bibinfo{title}{Narrow-linewidth iii-v/si/si3n4 laser using multilayer heterogeneous integration}.
\newblock \emph{\bibinfo{journal}{Optica}} \textbf{\bibinfo{volume}{7}}, \bibinfo{pages}{20--21} (\bibinfo{year}{2020}).

\bibitem{xiang2021high}
\bibinfo{author}{Xiang, C.} \emph{et~al.}
\newblock \bibinfo{title}{High-performance lasers for fully integrated silicon nitride photonics}.
\newblock \emph{\bibinfo{journal}{Nature communications}} \textbf{\bibinfo{volume}{12}}, \bibinfo{pages}{6650} (\bibinfo{year}{2021}).

\bibitem{zhang2018transfer}
\bibinfo{author}{Zhang, J.} \emph{et~al.}
\newblock \bibinfo{title}{Transfer-printing-based integration of a iii-v-on-silicon distributed feedback laser}.
\newblock \emph{\bibinfo{journal}{Optics express}} \textbf{\bibinfo{volume}{26}}, \bibinfo{pages}{8821--8830} (\bibinfo{year}{2018}).

\bibitem{zhang2023iii}
\bibinfo{author}{Zhang, J.} \emph{et~al.}
\newblock \bibinfo{title}{Iii-v-on-si dfb laser with co-integrated power amplifier realized using micro-transfer printing}.
\newblock \emph{\bibinfo{journal}{IEEE Photonics Technology Letters}} \textbf{\bibinfo{volume}{35}}, \bibinfo{pages}{593--596} (\bibinfo{year}{2023}).

\bibitem{roelkens2006laser}
\bibinfo{author}{Roelkens, G.}, \bibinfo{author}{Thourhout, D.~V.}, \bibinfo{author}{Baets, R.}, \bibinfo{author}{N{\"o}tzel, R.} \& \bibinfo{author}{Smit, M.}
\newblock \bibinfo{title}{Laser emission and photodetection in an inp/ingaasp layer integrated on and coupled to a silicon-on-insulator waveguide circuit}.
\newblock \emph{\bibinfo{journal}{Optics express}} \textbf{\bibinfo{volume}{14}}, \bibinfo{pages}{8154--8159} (\bibinfo{year}{2006}).

\bibitem{tran2022extending}
\bibinfo{author}{Tran, M.~A.} \emph{et~al.}
\newblock \bibinfo{title}{Extending the spectrum of fully integrated photonics to submicrometre wavelengths}.
\newblock \emph{\bibinfo{journal}{Nature}} \textbf{\bibinfo{volume}{610}}, \bibinfo{pages}{54--60} (\bibinfo{year}{2022}).

\bibitem{zhang2023photonic}
\bibinfo{author}{Zhang, Z.} \emph{et~al.}
\newblock \bibinfo{title}{Photonic integration platform for rubidium sensors and beyond}.
\newblock \emph{\bibinfo{journal}{Optica}} \textbf{\bibinfo{volume}{10}}, \bibinfo{pages}{752--753} (\bibinfo{year}{2023}).

\bibitem{castro2025integrated}
\bibinfo{author}{Castro, J.~E.} \emph{et~al.}
\newblock \bibinfo{title}{Integrated mode-hop-free tunable lasers at 780 nm for chip-scale classical and quantum photonic applications}.
\newblock \emph{\bibinfo{journal}{APL Photonics}} \textbf{\bibinfo{volume}{10}}, \bibinfo{pages}{036102} (\bibinfo{year}{2025}).

\bibitem{fan2016optically}
\bibinfo{author}{Fan, Y.} \emph{et~al.}
\newblock \bibinfo{title}{Optically integrated inp--si $ \_3 $ n $ \_4 $ hybrid laser}.
\newblock \emph{\bibinfo{journal}{IEEE photonics journal}} \textbf{\bibinfo{volume}{8}}, \bibinfo{pages}{1--11} (\bibinfo{year}{2016}).

\bibitem{fan2020hybrid}
\bibinfo{author}{Fan, Y.} \emph{et~al.}
\newblock \bibinfo{title}{Hybrid integrated inp-si3n4 diode laser with a 40-hz intrinsic linewidth}.
\newblock \emph{\bibinfo{journal}{Optics express}} \textbf{\bibinfo{volume}{28}}, \bibinfo{pages}{21713--21728} (\bibinfo{year}{2020}).

\bibitem{winkler2024widely}
\bibinfo{author}{Winkler, L.~V.} \emph{et~al.}
\newblock \bibinfo{title}{Widely tunable and narrow-linewidth hybrid-integrated diode laser at 637 nm}.
\newblock \emph{\bibinfo{journal}{Optics express}} \textbf{\bibinfo{volume}{32}}, \bibinfo{pages}{29710--29720} (\bibinfo{year}{2024}).

\bibitem{boller2020hybrid}
\bibinfo{author}{Boller, K.-J.} \emph{et~al.}
\newblock \bibinfo{title}{Hybrid integrated semiconductor lasers with silicon nitride feedback circuits}.
\newblock In \emph{\bibinfo{booktitle}{Photonics}}, vol.~\bibinfo{volume}{7}, \bibinfo{pages}{4} (\bibinfo{organization}{Multidisciplinary Digital Publishing Institute}, \bibinfo{year}{2020}).

\bibitem{klaver2021self}
\bibinfo{author}{Klaver, Y.}, \bibinfo{author}{Epping, J.~P.}, \bibinfo{author}{Roeloffzen, C.~G.} \& \bibinfo{author}{Marpaung, D.~A.}
\newblock \bibinfo{title}{Self-mode-locking in a high-power hybrid silicon nitride integrated laser}.
\newblock \emph{\bibinfo{journal}{Optics letters}} \textbf{\bibinfo{volume}{47}}, \bibinfo{pages}{198--201} (\bibinfo{year}{2021}).

\bibitem{winkler2024chip}
\bibinfo{author}{Winkler, L.~V.} \emph{et~al.}
\newblock \bibinfo{title}{Chip-integrated extended-cavity mode-locked laser in the visible}.
\newblock \emph{\bibinfo{journal}{Optics letters}} \textbf{\bibinfo{volume}{49}}, \bibinfo{pages}{6916--6919} (\bibinfo{year}{2024}).

\bibitem{vissers2021hybrid}
\bibinfo{author}{Vissers, E.}, \bibinfo{author}{Poelman, S.}, \bibinfo{author}{de~Beeck, C.~O.}, \bibinfo{author}{Van~Gasse, K.} \& \bibinfo{author}{Kuyken, B.}
\newblock \bibinfo{title}{Hybrid integrated mode-locked laser diodes with a silicon nitride extended cavity}.
\newblock \emph{\bibinfo{journal}{Optics Express}} \textbf{\bibinfo{volume}{29}}, \bibinfo{pages}{15013--15022} (\bibinfo{year}{2021}).

\bibitem{vissers2022hybrid}
\bibinfo{author}{Vissers, E.} \emph{et~al.}
\newblock \bibinfo{title}{Hybrid integrated mode-locked laser using a gaas-based 1064 nm gain chip and a sin external cavity}.
\newblock \emph{\bibinfo{journal}{Optics Express}} \textbf{\bibinfo{volume}{30}}, \bibinfo{pages}{42394--42405} (\bibinfo{year}{2022}).

\bibitem{roelkens2024present}
\bibinfo{author}{Roelkens, G.} \emph{et~al.}
\newblock \bibinfo{title}{Present and future of micro-transfer printing for heterogeneous photonic integrated circuits}.
\newblock \emph{\bibinfo{journal}{APL Photonics}} \textbf{\bibinfo{volume}{9}} (\bibinfo{year}{2024}).

\bibitem{justice2012wafer}
\bibinfo{author}{Justice, J.} \emph{et~al.}
\newblock \bibinfo{title}{Wafer-scale integration of group iii--v lasers on silicon using transfer printing of epitaxial layers}.
\newblock \emph{\bibinfo{journal}{Nature Photonics}} \textbf{\bibinfo{volume}{6}}, \bibinfo{pages}{610--614} (\bibinfo{year}{2012}).

\bibitem{loi2018thermal}
\bibinfo{author}{Loi, R.} \emph{et~al.}
\newblock \bibinfo{title}{Thermal analysis of inp lasers transfer printed to silicon photonics substrates}.
\newblock \emph{\bibinfo{journal}{Journal of lightwave technology}} \textbf{\bibinfo{volume}{36}}, \bibinfo{pages}{5935--5941} (\bibinfo{year}{2018}).

\bibitem{juvert2018integration}
\bibinfo{author}{Juvert, J.} \emph{et~al.}
\newblock \bibinfo{title}{Integration of etched facet, electrically pumped, c-band fabry-p{\'e}rot lasers on a silicon photonic integrated circuit by transfer printing}.
\newblock \emph{\bibinfo{journal}{Optics express}} \textbf{\bibinfo{volume}{26}}, \bibinfo{pages}{21443--21454} (\bibinfo{year}{2018}).

\bibitem{uzun2023integration}
\bibinfo{author}{Uzun, A.} \emph{et~al.}
\newblock \bibinfo{title}{Integration of edge-emitting quantum dot lasers with different waveguide platforms using micro-transfer printing}.
\newblock \emph{\bibinfo{journal}{IEEE Journal of Selected Topics in Quantum Electronics}} \textbf{\bibinfo{volume}{29}}, \bibinfo{pages}{1--10} (\bibinfo{year}{2023}).

\bibitem{coddington2016dual}
\bibinfo{author}{Coddington, I.}, \bibinfo{author}{Newbury, N.} \& \bibinfo{author}{Swann, W.}
\newblock \bibinfo{title}{Dual-comb spectroscopy}.
\newblock \emph{\bibinfo{journal}{Optica}} \textbf{\bibinfo{volume}{3}}, \bibinfo{pages}{414--426} (\bibinfo{year}{2016}).

\bibitem{mcferran2005low}
\bibinfo{author}{McFerran, J.~J.} \emph{et~al.}
\newblock \bibinfo{title}{Low-noise synthesis of microwave signals from an optical source}.
\newblock \emph{\bibinfo{journal}{Electronics Letters}} \textbf{\bibinfo{volume}{41}}, \bibinfo{pages}{650--651} (\bibinfo{year}{2005}).

\bibitem{telle1999carrier}
\bibinfo{author}{Telle, H.~R.} \emph{et~al.}
\newblock \bibinfo{title}{Carrier-envelope offset phase control: A novel concept for absolute optical frequency measurement and ultrashort pulse generation}.
\newblock \emph{\bibinfo{journal}{Applied Physics B}} \textbf{\bibinfo{volume}{69}}, \bibinfo{pages}{327--332} (\bibinfo{year}{1999}).

\bibitem{marpaung2019integrated}
\bibinfo{author}{Marpaung, D.}, \bibinfo{author}{Yao, J.} \& \bibinfo{author}{Capmany, J.}
\newblock \bibinfo{title}{Integrated microwave photonics}.
\newblock \emph{\bibinfo{journal}{Nature photonics}} \textbf{\bibinfo{volume}{13}}, \bibinfo{pages}{80--90} (\bibinfo{year}{2019}).

\bibitem{keller2021ultrafast}
\bibinfo{author}{Keller, U.} \& \bibinfo{author}{Paschotta, R.}
\newblock \emph{\bibinfo{title}{Ultrafast lasers}} (\bibinfo{publisher}{Springer}, \bibinfo{year}{2021}).

\bibitem{571743}
\bibinfo{author}{Keller, U.} \emph{et~al.}
\newblock \bibinfo{title}{Semiconductor saturable absorber mirrors (sesam's) for femtosecond to nanosecond pulse generation in solid-state lasers}.
\newblock \emph{\bibinfo{journal}{IEEE Journal of Selected Topics in Quantum Electronics}} \textbf{\bibinfo{volume}{2}}, \bibinfo{pages}{435--453} (\bibinfo{year}{1996}).

\bibitem{paschotta2004noise}
\bibinfo{author}{Paschotta, R.}
\newblock \bibinfo{title}{Noise of mode-locked lasers (part ii): timing jitter and other fluctuations}.
\newblock \emph{\bibinfo{journal}{Applied Physics B}} \textbf{\bibinfo{volume}{79}}, \bibinfo{pages}{163--173} (\bibinfo{year}{2004}).

\bibitem{paschotta2006optical}
\bibinfo{author}{Paschotta, R.}, \bibinfo{author}{Schlatter, A.}, \bibinfo{author}{Zeller, S.}, \bibinfo{author}{Telle, H.} \& \bibinfo{author}{Keller, U.}
\newblock \bibinfo{title}{Optical phase noise and carrier-envelope offset noise of mode-locked lasers}.
\newblock \emph{\bibinfo{journal}{Applied Physics B}} \textbf{\bibinfo{volume}{82}}, \bibinfo{pages}{265--273} (\bibinfo{year}{2006}).

\bibitem{komljenovic2015widely}
\bibinfo{author}{Komljenovic, T.} \emph{et~al.}
\newblock \bibinfo{title}{Widely tunable narrow-linewidth monolithically integrated external-cavity semiconductor lasers}.
\newblock \emph{\bibinfo{journal}{IEEE Journal of Selected Topics in Quantum Electronics}} \textbf{\bibinfo{volume}{21}}, \bibinfo{pages}{214--222} (\bibinfo{year}{2015}).

\bibitem{van2020ring}
\bibinfo{author}{Van~Rees, A.} \emph{et~al.}
\newblock \bibinfo{title}{Ring resonator enhanced mode-hop-free wavelength tuning of an integrated extended-cavity laser}.
\newblock \emph{\bibinfo{journal}{Optics express}} \textbf{\bibinfo{volume}{28}}, \bibinfo{pages}{5669--5683} (\bibinfo{year}{2020}).

\bibitem{vanackere2023heterogeneous}
\bibinfo{author}{Vanackere, T.} \emph{et~al.}
\newblock \bibinfo{title}{Heterogeneous integration of a high-speed lithium niobate modulator on silicon nitride using micro-transfer printing}.
\newblock \emph{\bibinfo{journal}{APL Photonics}} \textbf{\bibinfo{volume}{8}} (\bibinfo{year}{2023}).

\bibitem{niels2025high}
\bibinfo{author}{Niels, M.} \emph{et~al.}
\newblock \bibinfo{title}{A high-speed heterogeneous lithium tantalate silicon photonics platform}.
\newblock \emph{\bibinfo{journal}{arXiv preprint arXiv:2503.10557}}  (\bibinfo{year}{2025}).

\bibitem{billet2022gallium}
\bibinfo{author}{Billet, M.} \emph{et~al.}
\newblock \bibinfo{title}{Gallium phosphide-on-insulator integrated photonic structures fabricated using micro-transfer printing}.
\newblock \emph{\bibinfo{journal}{Optical Materials Express}} \textbf{\bibinfo{volume}{12}}, \bibinfo{pages}{3731--3737} (\bibinfo{year}{2022}).

\end{thebibliography}

\begin{acknowledgments}
We acknowledge funding by the Horizon Europe programme of the European Union (Grant agreement ID: 101070622) and the Flemish Research Council (FWO PhD fellowship grants 1SF9322N and 11F8120N).\\
\end{acknowledgments}

\section*{Author contributions}
M.K., S.C., and B.K. conceived the idea of the project. M.K. and S.C. designed the III-V coupons. M.K. and S.C. fabricated the III-V coupons with assistance from M.B., K.A., and V.B.O. M.K. designed and simulated the SiN components and circuits and developed the integration technique. M.K. measured and characterized the lasers with assistance from K.V.G and B.K. G.J., S.S., M.R., P.V.D., R.J., and J.B. developed imec's 200 mm SiN photonics platform and provided the wafer. M.K. prepared figures and wrote the manuscript. All authors reviewed the manuscript. G.R., K.V.G., and B.K. supervised the project.

\section*{Competing interest statement}
The authors declare no competing interests.


\clearpage

\newgeometry{a4paper, left=1in, right=1in, top=1in, bottom=1in}  
\fontsize{10pt}{14pt}\selectfont  
\onecolumngrid 
\renewcommand{\thesection}{\arabic{section}} 
\renewcommand{\thepage}{S\arabic{page}} 

\setcounter{section}{0} 
\setcounter{page}{1} 
\setcounter{figure}{0}

\renewcommand{\figurename}{Supplementary Fig.}
\renewcommand{\thefigure}{S\arabic{figure}} 

\begin{center}
    {\LARGE \textbf{Supplementary Material}}\\[1cm]
\end{center}

\title[Supplementary information]{Supplementary information for:\\Micro-Transfer Printed Continuous-Wave and Mode-Locked Laser Integration at 800 nm on a Silicon Nitride Platform}
\author{Max~Kiewiet$^{1,2,*}$, Stijn Cuyvers$^{1,2}$, Maximilien Billet$^{1,2}$, Konstantinos Akritidis$^{1,2}$, Valeria Bonito Oliva$^{1,2}$, Gaudhaman Jeevanandam$^{2}$, 
Sandeep Saseendran$^{2}$, Manuel Reza$^{2}$, Pol Van Dorpe$^{2}$, Roelof Jansen$^{2}$, Joost Brouckaert$^{2}$, G\"unther~Roelkens$^{1,2}$, Kasper Van Gasse$^{1,2}$ and Bart~Kuyken$^{1,2,\dagger}$\\  
\vspace{+0.1 in}
\textit{\small{
$^1$Photonics Research Group, INTEC, Ghent University - imec, 9052 Ghent, Belgium\\
$^2$imec, Kapeldreef 75, 3001 Leuven, Belgium. \\
{\small $*$Max.Kiewiet@ugent.be, $^\dagger$Bart.Kuyken@ugent.be}}}}

\date{\today}

\maketitle

\section*{\label{sec:level1}Supplementary note 1: Detailed micro-transfer printing integration process flow}
In this section, we describe the integration process flow for micro-transfer printing in more detail. The integration process uses dies from the imec 200-mm wafer-scale silicon nitride process. While the integration process happens at die-level, it can readily be adapted to create a wafer-scale process. The process to create III-V laser coupons starts with epitaxially grown (Al)GaAs layers as shown in Supplementary Fig. \ref{fig:epistack}. In this layer stack, two vertical regions can be distinguished. The top region is a standard edge-emitting laser stack, featuring four quantum wells, which can be taken from any standard ridge-laser process. The region below it consists of layers whose thicknesses are tuned to vertically align the III-V mode to the SiN mode after MTP.

\begin{figure}[h!]
    \centering
    \includegraphics[]{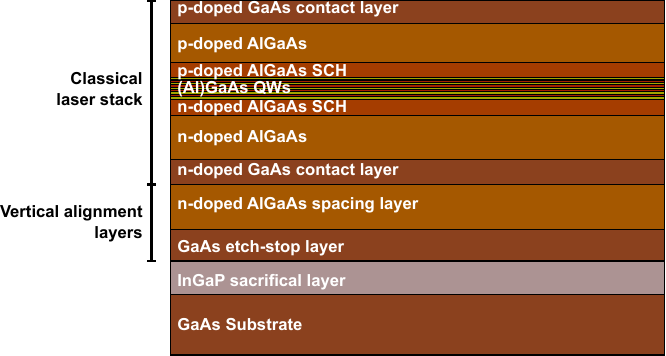}
    \caption{Epitaxial layer stack for butt-coupled micro-transfer printing.}
    \label{fig:epistack}
\end{figure}

This III-V material is subsequently processed to form laser coupons which are encapsulated in photoresist and under-etched to suspend them for pick-up. Next the integration flow as outlined in Supplementary Fig. \ref{fig:integration} is followed. This is described in more detail below:

\begin{figure}[h!]
    \centering
    \includegraphics[width=\textwidth]{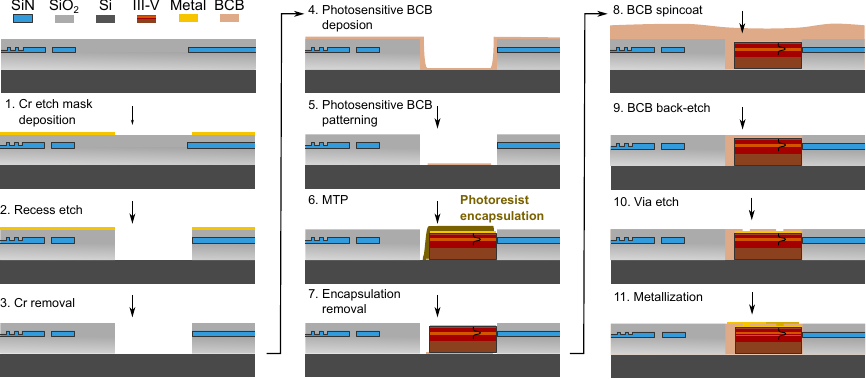}
    \caption{Detailed MTP integration process flow.}
    \label{fig:integration}
\end{figure}

\begin{enumerate}
    \item A chromium metal hard mask is deposited through lift-off.
    \item A recess is etched through the SiOx and SiNx layers using ICP-RIE until the silicon substrate or the desired etch depth is reached.
    \item The remaining Chromium metal is removed using Transene CR etchant 1020 and oxygen plasma.
    \item Photosensitive B-staged bisbenzocyclobutene (photo-BCB) Cyclotene 4000 series is diluted and spincoated to form an approximately 50 nm-thick layer. Due to the topography, photo-BCB will pool at the edges of recess (about 3-5 $\mathrm{\mu m}$), limiting the proximity that the laser facet can be printed from the SiN facet.
    \item Through UV-litho, the photo-BCB is patterned in a rectangle which is inset about 5 $\mathrm{\mu m}$ in the etched recess, removing any pooled-up photo-BCB.
    \item The III-V coupon is micro-transfer printed in the recess minimizing facet distance.
    \item The photoresist encapsulation is removed through a combination of an acetone rinse and oxygen plasma.
    \item A thick ($> 5\ \mathrm{\mu m}$) BCB (Cyclotene 3000 series) is spin coated and cured to planarize the sample for metallization and to fill the space between the III-V facet and the SiN facet to reduce reflections.
    \item The BCB is etched back to finish the planarization.
    \item Vias are etched in the BCB to access the III-V contact pads.
    \item Metal contact pads are deposited to finish the metallization.
\end{enumerate}
The steps following BCB deposition (8-10) can be significantly simplified in future by using photo-patternable BCB, eliminating the need for back-etching and via etching. 

\newpage
\section*{\label{sec:level1}Supplementary note 2: Direct butt-coupling approach}
\subsection{\label{sec:level2}Optimization of the SiN mode shape.}
In the direct butt-coupling approach, the III-V mode is directly aligned with the SiN waveguide mode. This allows for broad-band coupling without the need for intermediate taper structures. In this work, two SiN waveguide platforms are used: the 200 mm platform of imec, and an in-house electron-beam lithography (EBL) platform. For the 200 mm platform,  a shallow etch was used to match the shape of the SiN mode optimally to the III-V mode. For this, an inverted taper was used to transition from the unetched 300 nm SiN waveguide to a 60 nm-thick and 3000 nm-wide waveguide to reach a mode overlap of 96\%. The loss of this taper was characterized using a taper number sweep, which is shown in Supplementary Fig. \ref{fig:taperloss}.
\begin{figure}[H]
    \centering
    \includegraphics[width=0.5\linewidth]{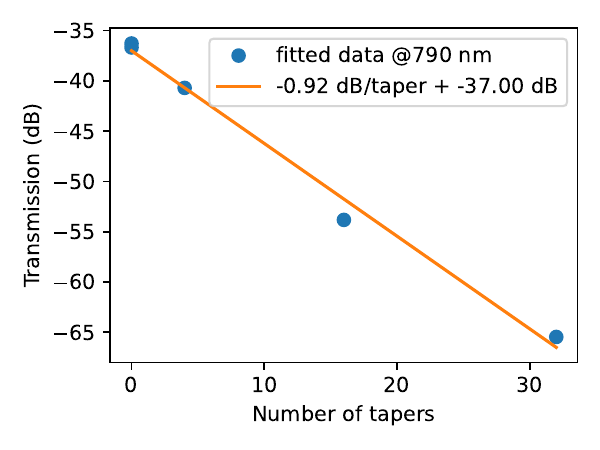}
    \caption{Taper loss sweep measurement along with linear fit.}
    \label{fig:taperloss}
\end{figure}
As the added coupling efficiency of the taper does not meaningfully exceed the insertion loss, for the EBL platform, the full 300 nm SiN layer was used, resulting in a SiN waveguide with a width of 3700 nm and simulated mode overlap of 84\%. The modes are compared in Supplementary Fig. \ref{fig:FDTDresults}a.

\subsection{\label{sec:level2}Simulations of the effect of misalignment on the coupling efficiency.}
To investigate the effect of misalignment of the micro-transfer printing process, Finite-Difference Time-Domain (FDTD) simulations were used which were carried out using Ansys Lumerical's FDTD suite. The simulations were set up as shown in Supplementary Fig. \ref{fig:FDTDresults}b. Two mode expansion monitors were used: port 1 excites the fundamental transverse-electric  (TE) mode in the III-V waveguide. Port 2 subsequently captures the transmission into the fundamental TE mode of the SiN waveguide. This is then calculated for a range of facet distances ($\Delta x$), lateral misalignment values ($\Delta y$), and vertical misalignment values ($\Delta z$). The parameter space is limited to estimated fabrication tolerances for $\Delta z$ and $\Delta x$ and estimated MTP alignment accuracy $\Delta y$. Typical field and refractive index profiles are shown in Supplementary Fig.~\ref{fig:FDTDresults}c for both the XY plane and the XZ plane. 

\begin{figure}[H]
    \centering
    \includegraphics[width=0.9\textwidth]{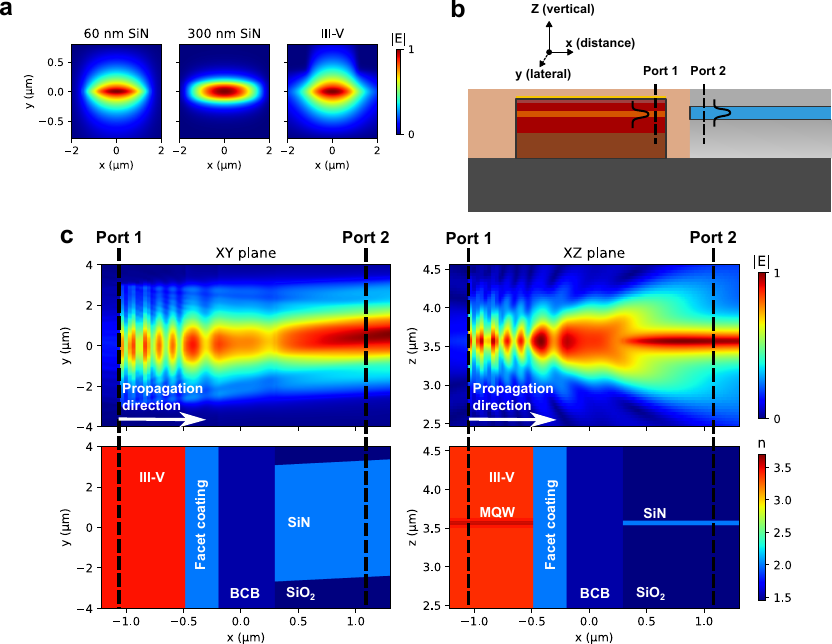}
    \caption{\textbf{FDTD simulation setup and profiles} a. Comparison of the 60 nm SiN mode, 300 nm SiN mode, and III-V waveguide mode. b. Illustration of the FDTD simulation setup. c. Typical electrical field amplitude and refractive index profiles for a facet distance of $\mathrm{0.5~\mu m}$ }
    \label{fig:FDTDresults}
\end{figure}

The result is shown in Supplementary Fig. \ref{fig:FDTDsweepl} for the 60 nm SiN mode and a III-V facet angle of 7 degrees. The lateral misalignment plotted is reported relative to the optimal placement of the III-V coupon, meaning that for larger facet distances, the coupon is moved along the propagation direction of the light in the BCB cladding layer. The vertical misalignment is plotted relative to the optimal height at 0 $\mathrm{\mu m}$ facet distance. The simulations show efficient coupling with a maximum of 86\% and more than 50\% within lateral fabrication tolerances of $\mathrm{\pm 1.0~\mu m~(3\sigma)}$ and vertical tolerances of $\mathrm{\pm 0.2~\mu m}$. With further optimization of the integration and a better anti-reflective (AR) facet coating, these values can be brought closer to the optimal value of 92\% with current modes and no reflection.

\begin{figure}[H]
    \centering
    \includegraphics[width=0.9\textwidth]{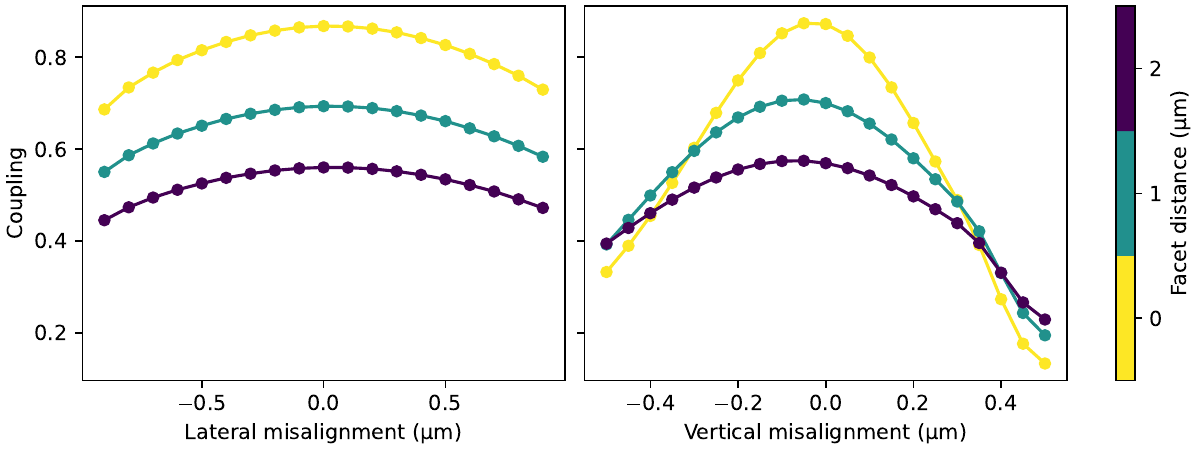}
    \caption{Simulated coupling values ($|S_{21}|^2$) for different 3D misalignment values.}
    \label{fig:FDTDsweepl}
\end{figure}

\subsection{\label{sec:level3}Simulations of the waveguide angle.}
To reduce reflection of RSOA coupons into the waveguide mode to prevent parasitic lasing of the RSOA independent of the feedback circuit, the front waveguide facet is angled. By increasing the facet angle, reflections into the waveguide mode are significantly decreased to $\mathrm{-70\ dB}$, as shown in the FDTD simulations in Supplementary Fig. \ref{fig:facetangle}.

\begin{figure}[H]
    \centering
    \includegraphics[width=0.5\textwidth]{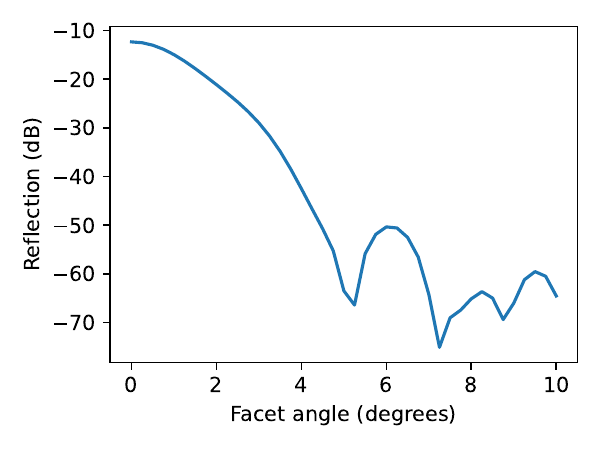}
    \caption{Reflection of the front III-V waveguide facet into the waveguide mode as a function of the facet angle.}
    \label{fig:facetangle}
\end{figure}

\newpage



\section*{\label{sec:level1}Supplementary note 3: Detailed (mode-locked) laser characterization}
In this note, we present additional characterization data for the 3.2 GHz mode-locked laser, as well as a comparison to other FSR mode-locked lasers in this work. Detailed mode-locking maps of the 3.2 GHz mode-locked laser are shown in Supplementary Fig. \ref{fig:modelockingmaps}.

\begin{figure}[H]
    \centering
    \includegraphics[width=\textwidth]{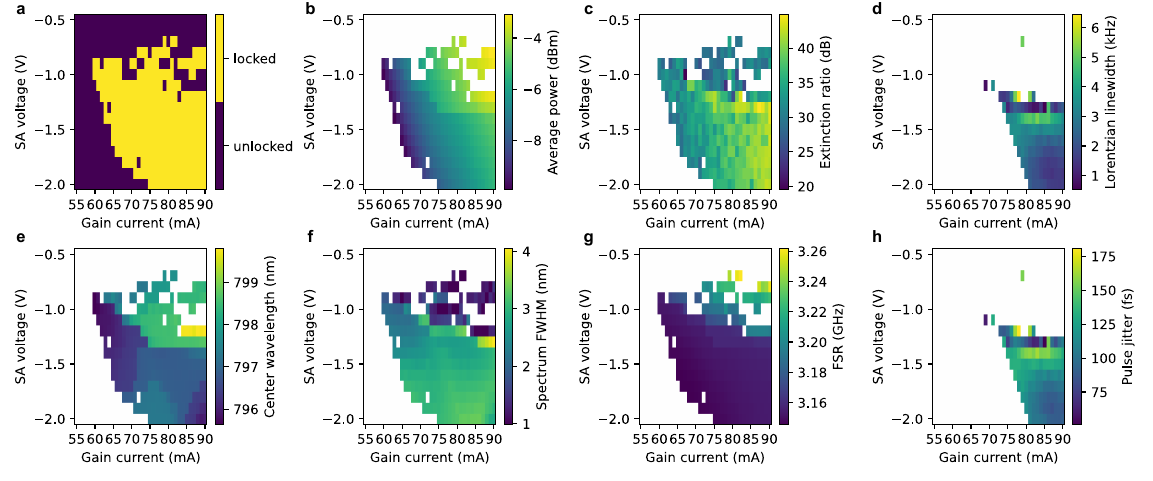}
    \caption{Mode-locking maps for a. Boolean mode-locking vs no mode-locking, b. Average laser power, c. Extinction ratio of the fundamental RF line, d. Lorentzian linewidth from phase noise measurement, where white data points mean no phase noise measurement was possible due to instable locking e. center wavelength of the optical spectrum, f. full width at half-maximum (FWHM) of the optical spectrum, g. free-spectral range of the mode-locking point, corresponding to the pulse repetition rate. h. Minimum pulse-to-pulse jitter as calculated from the Lorentzian linewidth.}
    \label{fig:modelockingmaps}
\end{figure}

In this work, a variety of mode-locked lasers with different FSR are characterized. A comparison between three mode-locked lasers is shown in Supplementary Fig. \ref{fig:mllcomparison}. Clearly, the most stable mode-locking with the widest optical spectrum is found for the 3.2 GHz FSR mode-locked laser. This is also the lowest average power mode-locked laser, as seen in the LIV characteristics in forward biased SA mode in Supplementary Fig. \ref{fig:forwardcomparison}. The difference in output power is most likely to be ascribed to the misalignment of the MTP process, affecting the III-V to SiN coupling.
\begin{figure}[H]
    \centering
    \includegraphics[width=0.5\linewidth]{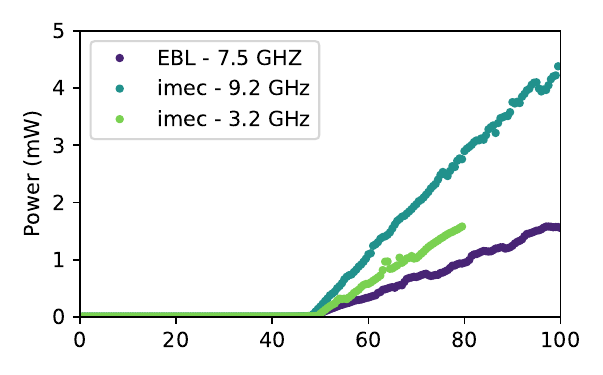}
    \caption{Waveguide-coupled output power of lasers in forward-biased SA (Fabry-Perot) mode.}
    \label{fig:forwardcomparison}
\end{figure}

\begin{figure}[H]
    \centering
    \includegraphics[]{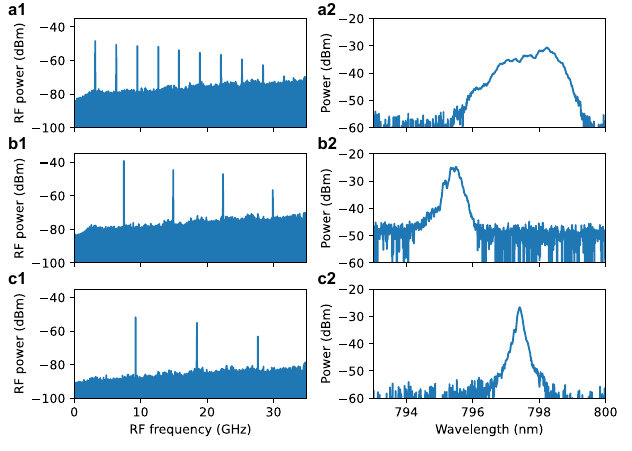}
    \caption{RF spectrum and optical spectrum of three different mode-locked lasers: a. 3.2 GHz FSR MLL on imec platform, b. 7.5 GHz FSR MLL on EBL platform, c. 9.2 GHz FSR MLL on imec platform.}
    \label{fig:mllcomparison}
\end{figure}

%

\end{document}